\documentclass[structabstract]{aa} 

\usepackage{amsmath, amssymb}
\usepackage{natbib}
\usepackage[english]{babel}
\usepackage{graphicx}
\usepackage{verbatim}
\usepackage{hyperref}
 \hypersetup{colorlinks=true,citecolor=black}
\begin{document}

\title{Constraints on off-axis jets from stellar tidal disruption flares}
\titlerunning{Off-axis tidal disruption jets}

\author{Sjoert van Velzen\inst{1} \and Dale A. Frail\inst{2} \and Elmar K\"ording\inst{1} \and Heino Falcke\inst{1,3,4} }

\institute{
  IMAPP, Radboud University, P.O. Box 9010, 6500 GL Nijmegen, The Netherlands \\ \email{s.vanvelzen@astro.ru.nl}
  \and National Radio Astronomy Observatory, Socorro, NM, USA
  \and ASTRON, Dwingeloo, The Netherlands 
  \and Max-Planck-Institut f\"ur Radioastronomie Bonn, Germany 
}

\date{Received 21 September 2012; Accepted 13 December 2012}

\abstract
{  Many decades of observations of active galactic nuclei (AGN) and X-ray binaries have shown that relativistic jets are ubiquitous when compact objects accrete. One could therefore anticipate the launch of a jet after a star is disrupted and accreted by a massive black hole. This birth of a relativistic jet may have been observed recently in two stellar tidal disruption flares (TDFs), which were discovered in gamma-rays by Swift. Yet no transient radio emission has been detected from the tens of TDF candidates that were discovered at optical to soft X-ray frequencies. Because the sample that was followed-up at radio frequencies is small, the non-detections can be explained by Doppler boosting, which reduces the jet flux for off-axis observers. And since the existing follow-up observation are mostly within $\sim 10$ months of the discovery, the non-detections can also be due to a delay of the radio emission with respect to the time of disruption.} 
{  We wish to test the conjecture that all TDFs launch jets. }
{We present 5 GHz follow-up observations with the Jansky~VLA of seven known TDFs, a significant increase of the number of radio observations of these events. To avoid missing  delayed jet emission, our observations probe 1--8 years since the estimated time of disruption. }
{None of the sources are detected, with very deep upper limits at the 10 micro Jansky level. These observations rule out the hypothesis that these TDFs launched jets similar to radio-loud quasars. We also constrain the possibility that the flares hosted a jet identical to Sw~1644+57, the first and best-sampled relativistic TDF. } 
{ We thus obtain evidence for a dichotomy in the stellar tidal disruption population, implying that the jet launching mechanism is sensitive to the parameters of the disruption. }\keywords{}
  \maketitle

\section{Introduction}
The disruption of a star by a massive black hole leads to arguably the most spectacular form of accretion onto these compact objects. 
The stellar debris that remains bound after the disruption returns to the black hole at a rate that initially can exceed the Eddington limit ($\dot{M}_{\rm Edd}$) by many orders of magnitude. This fallback rate declines with a power law index of $-5/3$ \citep{Rees88, Phinney89}, reaching 1\% of $\dot{M}_{\rm Edd}$ within a few to ten years. A tidal disruption flare (TDF) may thus be used to sample different modes of accretion \citep*[e.g.,][]{Abramowicz13} for a single supermassive black hole.
Considerable effort is needed to simulate the dynamics of the disruption \citep[e.g.,][]{Nolthenius82,EvansKochanek89,Rosswog09, Guillochon12} and to estimate the resulting optical to X-ray light curve of the flare \citep[e.g.,][]{loebUlmer97,bogdanovic04, strubbe_quataert09,Lodato11}. Efficient detection to obtain a large sample of TDFs is much anticipated, as this will allow, for example,  a study of the demographics of dormant black holes beyond the local universe \citep{Frank_Rees76, Lidskii_Ozernoi79}. 

Tens of (candidate) stellar tidal disruption events have been found by searching for flares in soft X-ray \citep[][]{KomossaBade99,Grupe99, KomossaGreiner99, Greiner00, Esquej08,Maksym10,Lin11,Saxton12}, UV \citep{Gezari06, Gezari08, Gezari09,Gezari12}, or optical surveys \citep{vanVelzen10,Drake11, Cenko12}, or based on spectra with extreme coronal lines \citep{Komossa08, Wang12}. None of these thermal flares are associated with a radio transient, but only a handful have been followed-up at this frequency.  The only tidal disruption candidates with a detected transient radio counterpart are those discovered in $\gamma$-rays by {\it Swift}: Sw~1644+57 \citep{Bloom11, Burrows11, Levan11, Zauderer11} and Sw~2058+05 \citep{Cenko12b}. Since the radio and X-ray emission of these two events most likely originates from a relativistic jetted outflow, they are often referred to as relativistic TDFs. In this paper we shall refer to the other class of TDFs as `thermal', since they are all discovered at optical to soft X-ray frequencies.

One is left to wonder why the two TDFs discovered with {\it Swift} are the only events with evidence for a newly-born jet. 
Interpreting this as a radio-loud/radio-quiet dichotomy, similar to the devision of radio-loudness in quasars \citep{Kellermann89, Falcke96, Sikora07}, would require that the tidal disruption jet launching mechanism is sensitive to the properties of the disruption (e.g., mass ratio, impact parameter, orientation of the orbit of the star with respect to the black hole spin, or circumnuclear environment). The explanation that quasars spend only a fraction of their time as radio-loud objects, similar to the jets in the `hard intermediate state' of X-ray binaries \citep[e.g.,][]{koerding06}, does not apply to tidal disruptions because their accretion rate is not constant. 
On the other hand, based on the observed fundamental plane of black hole accretion 
\citep*{Merloni03,Falcke04}, and the abundance of jets in low luminosity AGN \citep[][]{Nagar00} and X-ray binaries or microquasars \citep[][]{Mirabel99, Fender01}, one may postulate that \emph{all} stellar tidal disruptions launch jets. Likewise, \citet{Miller11} argue that the fundamental plane can be used to estimate the black hole mass of a TDF (if X-ray and radio observations of the flare are available).

If all stellar disruptions are indeed accompanied by a relativistic outflow, the current upper limits on the radio flux of the thermal TDFs can be explained by the orientation of this jet which can dramatically reduce the flux due to relativistic Doppler boosting. The current non-detections may also be explained by a delay of the radio emission of the jet with respect to the time of disruption. However, the number of TDFs that have been followed-up at radio frequencies is currently not sufficient to test this unification based on viewing angle. 

Recent advances in the hardware of the Very Large Array (VLA) have made it possible to obtain very deep radio observations of stellar tidal disruptions in a relatively short amount of time. To use this opportunity, we selected all thermal stellar tidal disruptions that occurred after 2004 for follow-up observations. 
These observations significantly increase the number of TDFs with deep radio observations. And because our radio observations span a wide range of times since the disruption, we can, for the first time, test the hypothesis that all stellar tidal disruptions launch jets. 

The remainder of paper is organized as follows. In sec. \ref{sec:models} we present two different tidal disruption (TD) jet models and compute off-axis light curves. In sec. \ref{sec:obs} we discuss the radio observations and sample selection. We use these observations to constrain the jet models in sec. \ref{sec:ana} and we close with a discussion in sec. \ref{sec:dis}.

\section{Tidal disruption jet models}\label{sec:models}
To be able to interpret our radio observations, we need a model that describes the radio emission of jets in accreting objects. In this section we therefore review two models of tidal disruption jets and we present off-axis light curves for these models. We have divided the models into two classes\footnote{Other models of TD jets \citep{Lei11, Krolik12, De-Colle12}, are not discussed here since these make no predictions for off-axis light curves at a given observer frequency.} based on the origin of the emitting particles: external or internal. In both models, some fraction of the accretion power ends up in the jet and the emission mechanism is synchrotron radiation.

\subsection{External model: off-axis light curves for Sw~1644+57}\label{sec:ex}

\begin{figure}
\begin{center}
\includegraphics[trim=0mm 0mm 0mm 6mm, clip, width=0.5\textwidth]{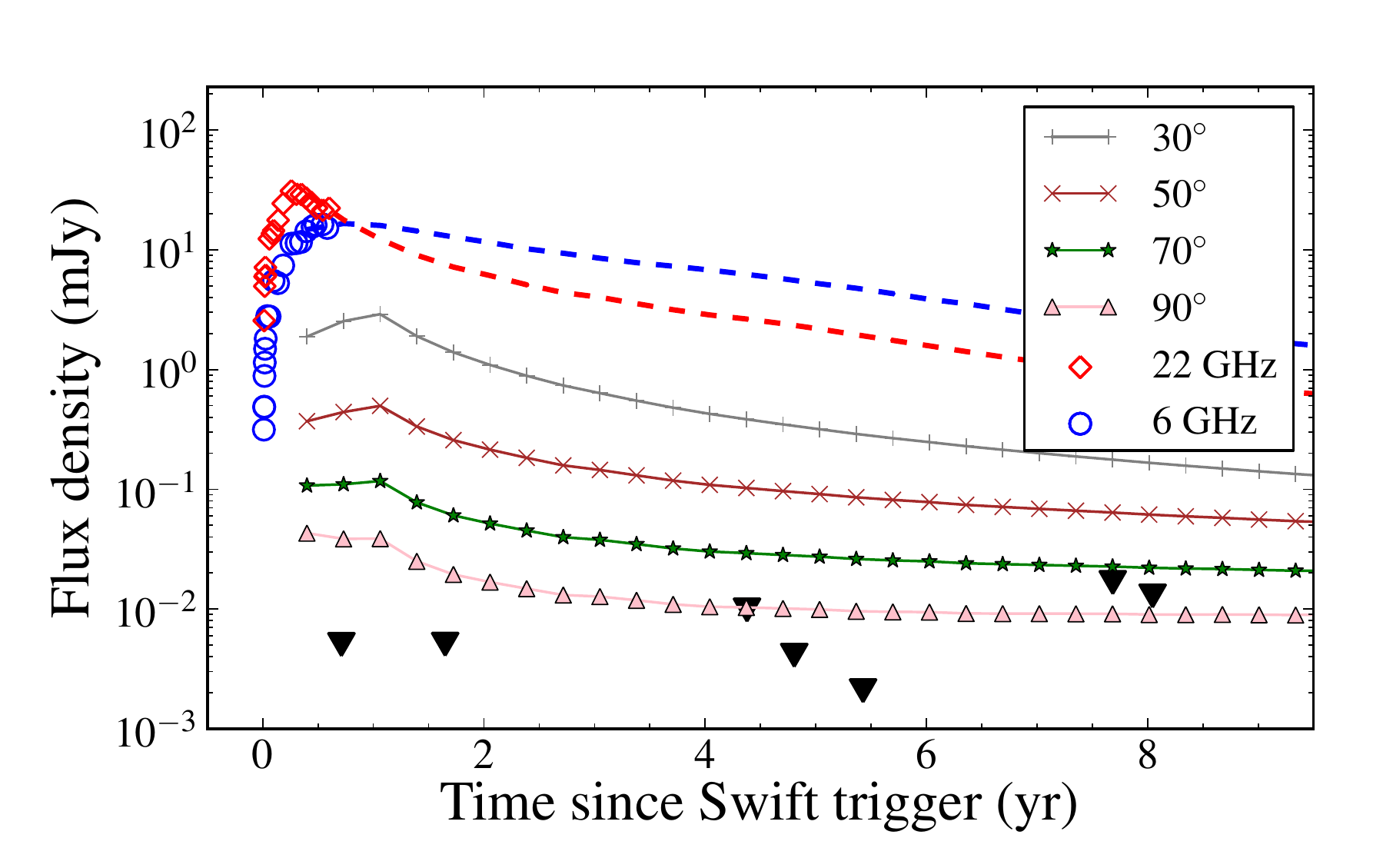} 
\caption{\small The observed light curve of Sw~1644+57 (open symbols), with the predicted late-time light curve (dashed lines) for a total jet energy of $E_j = 10^{52}~{\rm erg}$ \citep{Berger12}. We show the estimated 5~GHz light curve of different off-axis observers, Eq. \ref{eq:dshift}, assuming that the Lorentz factor of the jet decreases with $\Gamma_j\propto t^{-0.2}$, as inferred by \citet{Berger12}. We modified the extrapolated light curve to match a Sedov-Taylor solution, $L_j\propto t^{-1.9}$, when $\Gamma_j<2$. The 2-$\sigma$ upper limits on the radio flux of seven other TDF candidates (Table \ref{tab:can}) are shown with black triangles (we scaled these limits to the redshift of Sw~1644+57, see sec. \ref{sec:ana_vol}).}\label{fig:sw1644}
\end{center}
\end{figure}

The external model of radio emission from TD jets was first presented by \citet{Giannios11} and further developed by \citet*{Metzger12}. Shock interaction between the jet and the gaseous circumnuclear medium powers the emission, similar to afterglow models of gamma-ray bursts \citep[e.g.,][]{Sari95}.

The external model has been applied to the radio light curve of the relativistic TDF Sw~1644+57 \citep{Metzger12, Berger12}. We show the fit and predicted late-time light curve in Fig. \ref{fig:sw1644}. We note that this fit requires a continuous increase of the isotropic jet power during the first year of observations. 
 
The scaling of the synchrotron peak and self absorption frequency in the \citet{Metzger12} model of Sw~1644+57 are based on spherical expansion of an ultra-relativistic shell and thus require $\theta_j\Gamma_j<1$ (with $\theta_j$, $\Gamma_j$ the jet opening angle and Lorentz factor, respectively), plus an on-axis observer $i_{1}<1/\Gamma_j(t=0)$; both requirements are supported by the observed radio light curve \citep{Metzger12}. 

To compute the light curve for an off-axis observer, we first boost the observed on-axis flux $F_1(\nu)$ into the jet rest-frame
\begin{align}
L_j(\nu) = d_L^2 \delta_1^{3-\alpha} F_1(\nu) 
\end{align}
\citep[e.g.,][]{LindBlandford85,Jester08}. Here we introduced the Doppler factor for the on-axis observer $\delta_1=[\Gamma_j(1-\beta_j\cos i_1)]^{-1}$ with $\beta_j=v_j/c$, $\alpha$ is the spectral index defined as $F(\nu) \propto \nu^{\alpha}$, and $d_L$ is the luminosity distance. Next, we transform the jet luminosity to the off-axis observer using a different Doppler factor, $\delta_2$. If the size of the emitting region is small compared to the distance to the black hole, the time delay due to the geometrical separation of the synchrotron peak with frequency can be ignored, and we can estimate the flux for an observer sitting at $i_2$: 
\begin{align}\label{eq:dshift}
F_2(t, \nu) &= \left(\frac{\delta_2}{\delta_1} \right)^{3-\alpha(t)} F_1(t, \nu) \notag \\ 
&\approx \left(\frac{1-\beta_j(t)}{1-\beta_j(t)\cos i_{2} }\right)^{3-\alpha(t)} F_1(t, \nu).
\end{align}
Here $t$ is measured in the observer-frame, $\alpha(t)$ is obtained from the light curve, and we used $\cos i_1 \approx 1-\Gamma_j^{-2}/2 \sim 1$ to simplify the equation. If Sw~1644+57 was indeed a relativistic outflow that we observed on-axis, the light curve for off-axis observers depends only $\Gamma_j(t)$ and $i_2$.  The latter is a free-parameter, which we shall constrain by our follow-up observations in sec. \ref{sec:ana}.

To obtain $\Gamma_j(t)$ and the light curve beyond the last published radio observation of Sw~1644+57 (about one year after the {\it Swift} trigger), we used the external model of TD jets \citep{Metzger12} applied to the Sw 1644+57 radio data by \citet{Berger12}. The off-axis light curve that is derived here may thus be viewed as a test for this model. We consider two scenarios. First, we set $\Gamma(t>1~{\rm yr})=2$, and use the extrapolated light curve presented in \citet{Berger12}. This constant Lorentz factor is required because the extrapolated light curve is no longer valid when the jet slows down to mildly relativistic speed and lateral expansion becomes important. This happens at $\Gamma_j \lesssim 2$ \citep{Zhang09}. We also consider a decreasing jet velocity $\Gamma_j\propto t^{-0.2}$ (as inferred for Sw~1644+57), but modify the extrapolated light curve to match the non-relativistic Sedov-Taylor evolution when $\Gamma_j<2$. To estimate the light curve decay in the Sedov-Taylor phase, we assume an electron power-law index $p=2.5$ and describe the density of circumnuclear gas with a power-law of index $k=3/2$ \citep{Berger12} to find  $L_j\propto t^{-1.9}$ \citep[][]{Granot99,Leventis12}. We show the light curves in Fig.~\ref{fig:sw1644}.

\subsection{Internal model: off-axis light curves for known TDFs}\label{sec:intern}

\begin{figure}
\begin{center}
\includegraphics[trim=0mm 0mm 0mm 6mm, clip, width=0.5\textwidth]{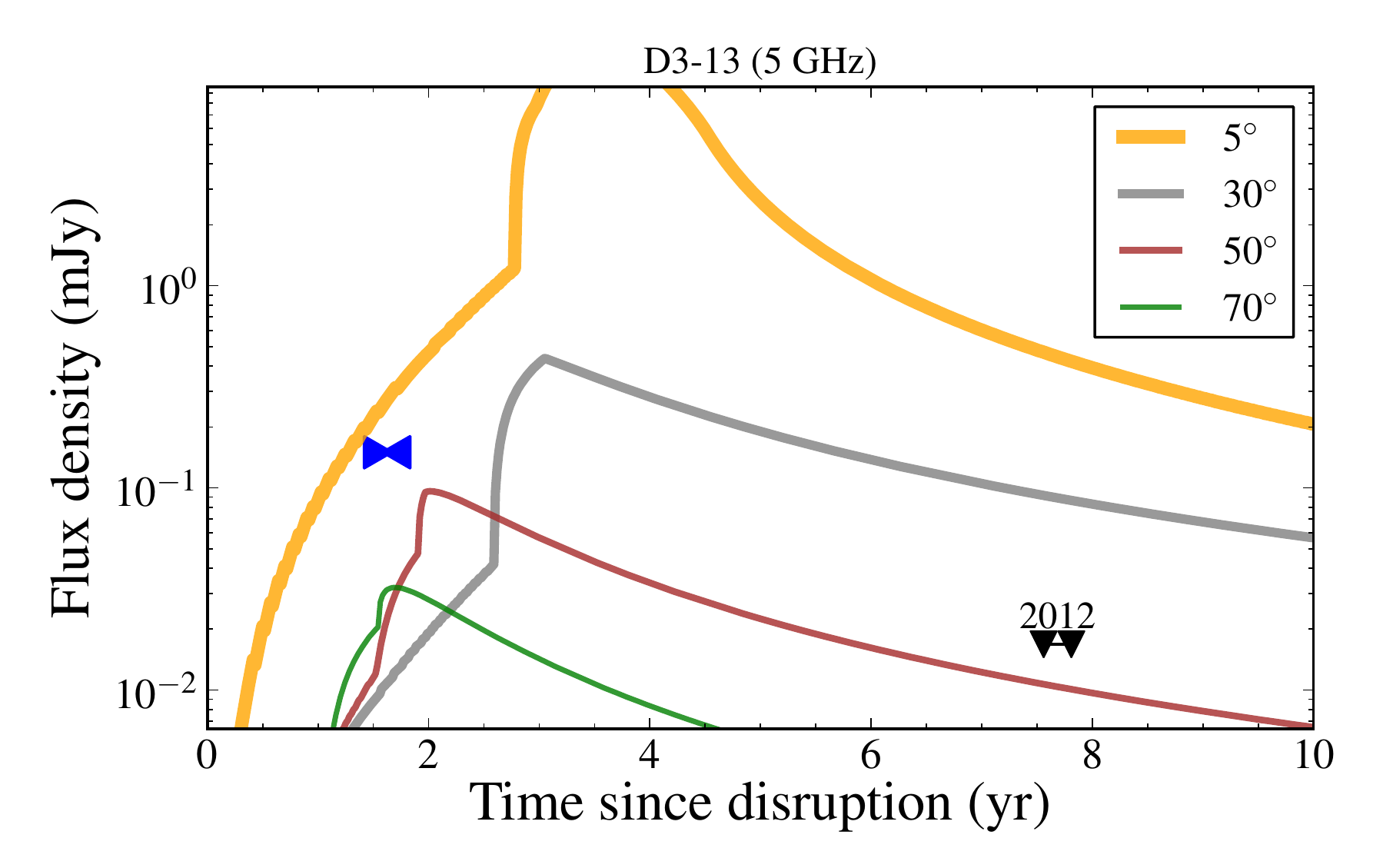} 
\caption{\small Predicted 5 GHz curves for the tidal disruption candidate D3-13 \citep{Gezari09} for the internal jet model in the optimistic scenario (Eq. \ref{eq:scen}$a$). The existing upper limits on the 1.4~GHz flux density \citep{Bower11} are shown with blue triangles, pointing left and right to indicate the uncertainty on the time of disruption. The downward pointing triangles labeled ``2012'' show our upper limits to the 5~GHz flux. For $\cos i <{\beta_j}$, the peak of the light curve decreases in time and magnitude as $i$ increases, because a fixed observed frequency corresponds to a smaller radius where the jet becomes optically thin to synchrotron self-absorption (i.e., $z_{\rm ssa}\propto \delta/\nu$). For $\cos i >{\beta_j}$ the light curve is compressed due to time retardation.}\label{fig:intern}
\end{center}
\end{figure}
The internal model of radio emission from TD jets was first presented in \citet*{vanVelzen10b} and further developed by \citet*{vanVelzen11}. The model is based on the simple idea of jet-disk coupling \citep{Rawlings91, Falcke95I}: a constant fraction of the accretion luminosity ($L_d$) is fed into the jet, $Q_j=q_j L_d$. The conversion from jet power ($Q_j$) to radio luminosity ($L_j$) follows by assuming equipartition between the energy in relativistic particles and magnetic fields, and has been calibrated using observations of AGN \citep{Falcke95II,Willott99,Koerding08}.

Stellar mass black holes show rapid switches from radio-loud ($q_j=0.2$) to radio-quiet ($q_j<0.002$) coupling as the accretion rate increases from sub-Eddington  to (near) the Eddington limit \citep*{Fender04}. 
Motivated by the growing evidence that accretion onto super-massive black holes can also be divided into these two modes \citep[e.g.,][]{Ho99, Ghisellini01,Falcke04,koerding06, Best12, Plotkin12}, we considered the following three scenarios for the jet-disk coupling in tidal disruptions:
\begin{align}\label{eq:scen}
  q_j = \left\{
  \begin{array}{l l c}
    0.2& \quad {\rm all~times} & (a) \\
    0.002& \quad \dot{M}(t)>2\% \dot{M}_{\rm Edd} &(b)\\
    0.2&\quad t<t_{\rm fallback} & (c) \\
      \end{array} 
    \right.
\end{align}
where each scenario reverts to the preceding one if the condition on $t$ or $\dot{M}$ is not true (e.g., $q_j=0.2$ when $\dot{M}<2\%\dot{M}_{\rm Edd}$ in all three scenarios). In the optimistic scenario ($a$), the TD jet behaves like a radio-loud quasar at all times. In the most conservative scenario ($b$), the jet becomes radio-loud only when the accretion drops below $<2\% \dot{M}_{\rm Edd}$ \citep{Maccarone03}. In scenario $c$, the system starts with a radio-loud burst during the onset of the accretion. We consider $c$ the most realistic scenario, since it most closely resembles the observed behavior of X-ray binaries. 

Besides $q_j$ (Eq. \ref{eq:scen}) the internal model requires a jet Lorentz factor and the disk luminosity as a function of time. The latter is obtained from the fallback rate of the stellar debris for a pericenter passage at the disruption radius, capped at the Eddington limit. We vary $\Gamma_j$ between 5, the default value in \citet{vanVelzen11}, and $\Gamma_j=2$. In Fig. \ref{fig:intern} we show an example light curve of the TDF candidate D3-13 \citep{Gezari09} for $\Gamma_j=5$, using the black hole mass as estimated from the luminosity of the host. 

In the internal TD jet model, the typical time scale of the light curve is set by the radius where the jet becomes optically thin to synchrotron self-absorption, $z_{\rm ssa}\propto \delta \nu^{-1} L_d^{2/3}$. This is different from our off-axis version of the external model, where we assumed that the emission is dominated by the head of the jet, which explains the dissimilar scaling of the peak of the light curve with the Doppler factor for the internal and external model.

\section{Observations}\label{sec:obs}
\begin{table}
\caption{\label{tab:prev} \small  Existing radio follow-up observations of TDF candidates that were discovered at optical to soft X-ray frequencies (our new observations are shown in Table~\ref{tab:can}).}.
\centering
\begin{tabular}{lcccc}
\hline \hline
name & $t_D$  & $F_\nu$ & $\nu$ & $\Delta t$ \\
         & (yr) & (mJy) & (GHz) & (yr) \\
\hline 
NGC 5905$^1$ & 1990.5 & $<0.15$ & 8.5 & 6.0 \\ 
D3$-$13$^2$  & 2004.5 & $<0.15$ & 1.4 & 1.8 \\
TDE2$^3$  & 2007.8 & $<0.10$ & 8.4 & 1.1 \\
CSS100217$^4$ & 2010.2 & $0.50\pm 0.03$ & 7.9 & 0.3 \\
SDSS J1201+30$^5$  & 2010.4 & $<0.22$ & 4.8& 1.4 \\
\hline 
RX J1624+7554$^6$ & 1990.8 & $<0.085$ & 3.0 & 21.8 \\
IC 3599$^7$ & 1990.9 & $0.19 \pm 0.03$ &  3.0 & 21.6 \\
RX J1420+5334$^8$ & 1990.9 & $ 0.11 \pm 0.02$ & 3.0 & 21.6 \\
RX J1242$-$1119$^9$ & 1992.5 & $<0.090$ & 3.0 & 20.0 \\
SDSS J1323+48$^{10}$ & 2003.9 & $<0.170$ & 3.0 & 8.6 \\
SDSS J1311$-$01$^{11}$ & 2004.1 &  $<0.095 $ & 3.0 & 8.4\\
\hline
\end{tabular}
\tablefoot{\small In the third column we show 5-$\sigma$ upper limits on the radio flux or the detected flux and 1-$\sigma$ uncertainty. $\Delta t$ denotes the time of the radio observation with respect to the estimated time of disruption ($t_D$). The radio observation of the first five candidates were published before mid-2012, the last six are taken from \citet{Bower13}.
We note that most of the TDF candidates that were discovered in the nineties also have post disruption radio upper limits from large radio surveys \citep[e.g., NVSS,][]{Condon98}, these are listed in \citet{Komossa02}.}
\tablebib{(1)~\citet{Bade96, Komossa02}, (2)~\citet{Gezari08, Bower11}, (3)~\citet{vanVelzen10}, (4)~\citet{Drake11}, (5)~\citet{Saxton12}, (6)~\citet{Grupe99}, (7)~\citet{KomossaBade99}, (8)~\citet{KomossaGreiner99}, (9)~\citet{KomossaGreiner99}, (10)~\citet{Esquej07}, (11)~\citet{Maksym10}. }
\end{table}
In Table \ref{tab:prev} we summarize the published radio follow-up observations of TDF candidates that were discovered at optical to soft X-ray wavelengths. To increase this sample, we selected all TDF candidates with an estimated time of disruption after 2004 for follow-up observations. This limit is used since the internal model of TD jets is no longer valid when the jet slows down significantly. Our sample also includes TDFs with existing radio upper limits, since the radio emission can be delayed with respect to the time of disruption. We removed CSS100217 from the sample because it is detected at 1~GHz \emph{before} the time of disruption  with a flat spectral index, indicating an AGN origin for the radio emission \citep{Drake11}. SDSS~J1201+30 was not selected for follow-up observations because this TDF was published after our observations were scheduled. Details of the data reduction of the remaining six candidates are summarized below.

The radio observations were carried out on the Karl G. Jansky Very
Large Array on 29 January 2012 under program 12A-005. We
observed at a central frequency of 5.0 GHz with 16 subbands each with
64 2 MHz channels, spanning 2 GHz of total bandwidth. The VLA was in
the C configuration yielding typical angular resolution of 4 arcsec.
The total observing time was 2.5 hrs, with integration times for
individual TDF sources varied from 18-30 min. Phase calibration was
carried out by making short observations of nearby point source
calibrators every 10 minutes, while amplitude and bandpass calibration
was achieved using an observation of 3C\,286 or 3C\,48 at the
beginning or end of each observing run. The data were reduced
following standard practice in the Astronomical Image Processing
System (AIPS) software package.

In addition to these data, we identified one public data set from the
VLA archive (project AS\thinspace{1020}) for the TDF candidate
PS1-10jh \citep{Gezari12}. These observation were made with the VLA on
29 March 2011 in the B configuration with two subbands (each with 64 2
MHz channels) centered at 4.83 and 4.96 GHz, for a total bandwidth of
256 MHz. The calibration and imaging of these data was similar to the method 
described above.

Our final sample that we shall use to constrain TD jet models thus consists of seven TDF candidates that were observed with the Jansky~VLA. We summarize the results of these observations in Table \ref{tab:can}. 

\begin{table}
\caption{\label{tab:can} \small Jansky~VLA observations at 5~GHz of TDF candidates discovered at UV or optical frequencies.}
\centering
\begin{tabular}{cccccc}
\hline \hline
name  & redshift & $M_{\rm BH}$ & $t_{\rm int}$ & $\sigma(F_\nu)$ & $\Delta t$ \\
         &   & $M_\odot\times 10^{7}$ & (min)                 & ($\mu$Jy)        & (yr) \\
\hline 
D1-9$^1$      & 0.326 & 5  & 30 & 9   & 8.0 \\
D3-13$^1$   & 0.370 & 2 & 18 & 8   & 7.6 \\
TDE1$^2$      & 0.136 &1 & 28 & 10 & 5.4 \\
D23H-1$^3$  & 0.186& 5 & 28 &  8  & 4.8  \\
TDE2$^2$    & 0.252& 5 & 25 & 12 & 4.3 \\
PTF10iya$^4$ & 0.224& 1 & 18 & 8  & 1.6 \\
PS1-10jh$^5$ & 0.170 & 0.4 & 39  & 15 & 0.71 \\
\hline
\end{tabular}
\tablefoot{We list the redshift and estimated black hole mass of these candidates in the second and third column, respectively. No significant emission was detected at the phase center of the images. We list integration time after removal of interference, the rms of the images, and the time of the observations with respect to the estimated time of disruption.}
\tablebib{(1)~\citet{Gezari08}, (2)~\citet{vanVelzen10}, (3)~\citet{Gezari09}, (4)~\citet{Cenko12}, (5)~\citet{Gezari12}.}
\end{table}

\section{Analysis}\label{sec:ana}
In this section we first compute the constraints that can be placed on TD jet models using our Jansky~VLA follow-up observation and then consider the potential of radio transient surveys. 

\subsection{Constraints from follow-up observation}\label{sec:ana_vol}
If we assume that the angle between the observer and the jet is drawn from a uniform distribution (on a sphere), we can calculate the probability of non-detections for a given flux density limit. One simply has to find the largest angle for which the predicted flux is above the flux limit and then calculate the probability to observe a jet within this angle. The flux limit is set at twice the rms of the radio image of each TDF. (This is lower than the limit for a blind-detection experiment since we use the threshold to find the probability of a non-detection, not to claim a discovery.) In Table \ref{tab:prop} we list the results of this exercise.

\begin{figure}[t]
\begin{center}
\includegraphics[trim=0mm 0mm 0mm 6mm, clip, width=0.5\textwidth]{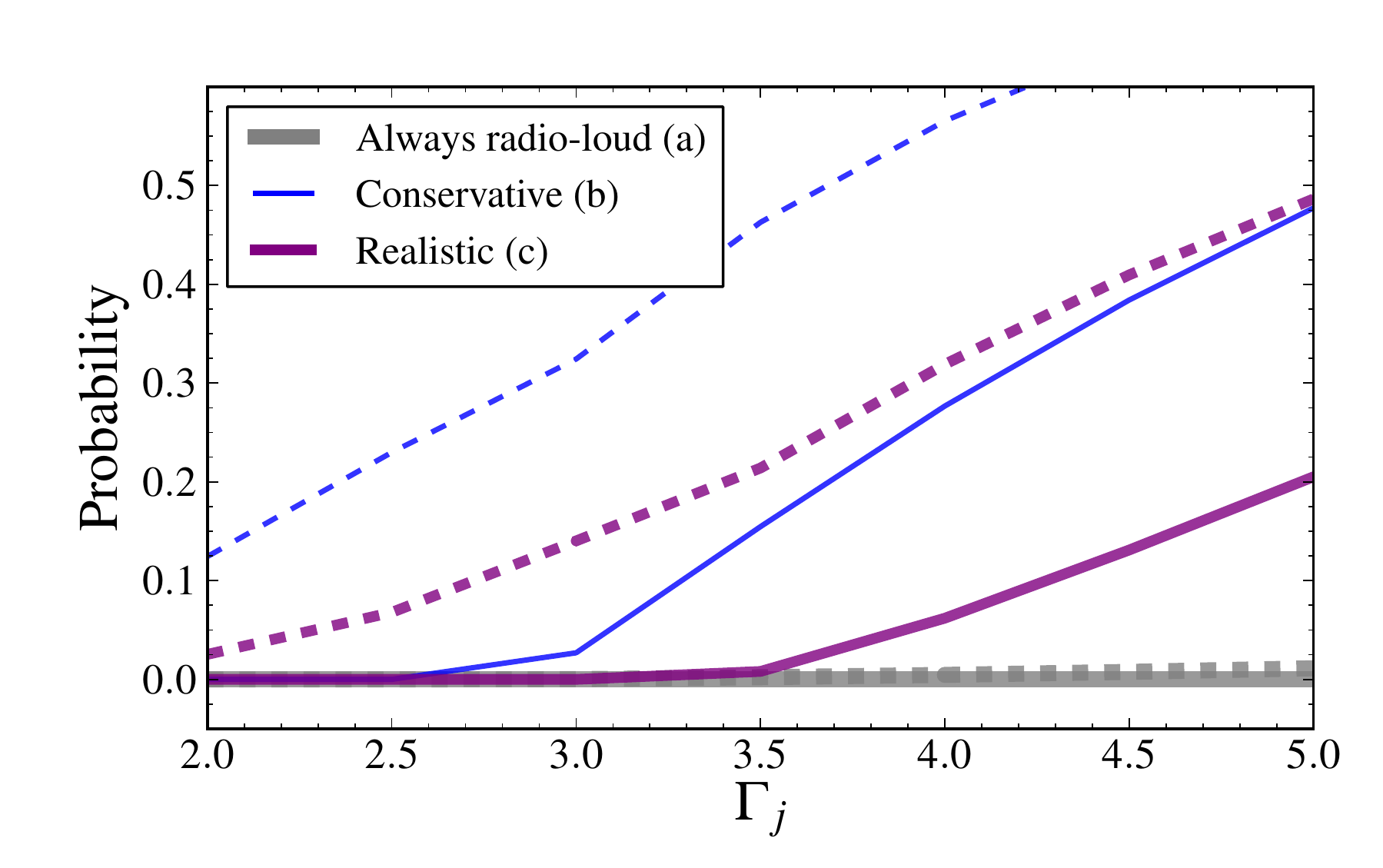} 
\caption{\small The probability of our data (i.e., no flux above the two times the image rms) for the three scenarios of the internal model (Eq. \ref{eq:scen}). The solid lines show $P_{7}$ the probability that all seven TDFs we observed indeed hosted a jet; the dashed lines show $P_{\ge 1}$ the probability that at least one flare hosted a jet. For $\Gamma_j<3$ the former hypothesis is ruled out at 95\% confidence for all scenarios. }\label{fig:Pgamma}
\end{center}
\end{figure}

The probability that all seven TDFs in our sample hosted jets, but were not detected due to Doppler boosting is $P_{7}=\Pi_i P_i $, with $P_i$ being the probability of the observations of each TDF candidate, as listed in Table \ref{tab:prop}. We also consider the possibility that, given our observations, \emph{at least one} of the seven TDFs hosted a jet, $P_{\ge 1}$. This is obtained by taking the mean value of the product of all combinations of the seven $P_i$'s (e.g., the probability that only one jet was launched is $P_{1}=\sum_i P_i /7$). For the optimistic scenario of the internal model (sec. \ref{sec:intern}) with $\Gamma_j=5$, four of the seven TDF candidates (D23H-1, TDE2, PTF10iya, PS1-10jh) should have yielded a detection above the 2-$\sigma$ level, hence $P_{7}=0$, while $P_{\ge 1}$=2\%. The probability that all of the other three TDF candidates hosted a jet is 1.7\%. For the most conservative scenario (Eq. \ref{eq:scen}$b$),  $P_{7}=48\%$, while for the realistic scenario (Eq. \ref{eq:scen}$c$) this is lower at 21\%. In Fig. \ref{fig:Pgamma} we show $P_{7}$ and $P_{\ge 1}$ for lower Lorentz factors; at $\Gamma_j<3$, the hypothesis that all seven TDFs hosted a jet is ruled out at 95\% confidence for all three scenarios of the internal jet model.

\begin{table}
\caption{\label{tab:prop} \small Probability (\%) that the jet orientation is such that the predicted flux is below the 2$\sigma$-level of our 5~GHz observation.}
\centering
\begin{tabular}{cccccc}
\hline \hline 
name & \multicolumn{3}{c}{Internal jet model} & \multicolumn{2}{c}{Sw~1644+57, off-axis} \\ 
 & $a$ & $c$ &  $b$ &  $\Gamma_j=2$ & $\Gamma_j \propto t^{-0.2} $ \\
\hline
D1-9 & 39 & 78 & 83 & 49 & 17 \\
D3-13 & 62 & 89 & 91 & 52 & 26 \\
TDE1 & 7 & 92 & 100 & 0 & 0 \\
D23H-1 & 0 & 52 & 70 & 0 & 0 \\
TDE2 & 0 & 75 & 98 & 20 & 1 \\
PTF10iya & 0 & 86 & 95 & 0 & 0 \\
PS1-10jh & 0 & 95 & 97 & 0 & 0 \\
\hline
\end{tabular}
\tablefoot{\small Zero probability implies that the predicted flux is above the threshold even for $i_2=\pi/2$, while $P_i=100\%$ implies the data cannot constrain the model. In the second to fourth column we list the results for the internal jet model, for the optimistic to the conservative scenario (Eq. \ref{eq:scen}), for $\Gamma_j=5$. In Fig. \ref{fig:Pgamma} we show the results for lower Lorentz factors. In the fifth and sixth column we give the probably of detecting a jet that is identical to Sw~1644+57, but observed off-axis, using two different estimates of the light curve past the last available observation (see sec. \ref{sec:ex}). }
\end{table}

Our upper limits also constrain the possibility that a jet similar to Sw~1644+57 was launched after the disruption. To place the Jansky~VLA observations on the estimated off-axis light curve (Eq. \ref{eq:dshift}), we equate the time of disruption to the time of the {\it Swift} trigger and we scale the flux using $(d_{L,{\rm Sw}}/d_L)^2$, with $d_{L,{\rm Sw}}$ the luminosity distance of Sw~1644+57.   From Fig. \ref{fig:sw1644} we see that our upper limits on the radio flux of five TDFs (TDE1, D23H-1, PTF10iya, and PS1-10jh) are inconsistent with the estimated off-axis light curve of Sw~1644+57 for all viewing angles and both versions of the late-time evolution we considered in sec. \ref{sec:ex}.

\subsection{Constraints from (future) radio transients surveys}
A different method to test whether jets like Sw~1644+57 are common to stellar tidal disruptions is to compute the rate of these transients. The snapshot rate (or areal density) at a given flux density limit $F_{\nu,{\rm lim}}$ can be estimated directly from the Sw~1644+57 light curve:
\begin{align}
  R(F_{\nu,{\rm lim}}) \sim 8\times 10^{-3} \, \Gamma_j^{-2}\, \left(\frac{F_{\nu,{\rm Sw}}}{F_{\nu,{\rm lim}}}\right)^{3/2}  \notag \\ 
  \frac{\Delta T \dot{N}_{\rm TDJ}}{10^{-5}}\frac{\rho_{\rm BH}}{5\times 10^{-3}~{\rm Mpc}^{-3}} \; {\rm deg}^{-2} \quad . \label{eq:snap}
\end{align}
Here $\Delta T$ is the time in years that the flux of Sw~1644+57 is above $F_{\nu,{\rm Sw}}$, $\rho_{\rm BH}$ is the black hole density, and $\dot{N}_{TDJ}$ is the rate of stellar tidal disruptions with jets. If Sw~1644+57 was a typical stellar tidal disruption, this rate should be of the same order as the TDF rate inferred from soft X-ray \citep[][]{Donley02} or optical \citep{vanVelzenFarrar_EPJW12} surveys, i.e., $\dot{N}_{\rm TDJ}\sim10^{-5}\,{\rm yr}^{-1}$. 

The  5~GHz light curve of Sw~1644+57 implies $\Delta T \approx 1$~yr for $F_{\nu,{\rm Sw}}=20$~mJy. For $\Gamma_j=2$, we can thus obtain the snapshot rate for a radio variability survey with a threshold at 10~mJy: $R(10\,{\rm mJy})=5\times 10^{-3}\, {\rm deg}^{-2} $. 
This rate is close to the existing upper limits on the snapshot rate at 5~GHz \citep[e.g.,][]{Scott96, Bower07, Bower11b} -- see \citet{Frail12} for a review. Hence the observed light curve of Sw~1644+67 implies that near-future radio variability survey will either measure or constrain $\dot{N}_{\rm TDJ}$.

\section{Conclusion \& Discussion}\label{sec:dis}
We obtained upper limits at the $\sim 10~\mu{\rm Jy}$ level of the 5~GHz flux of seven stellar tidal disruptions events that were discovered with optical/UV imaging surveys. This is three orders of magnitude lower than the recently discovered TDFs with radio emission, suggesting that stellar tidal disruptions come in different flavors, ranging from radio-loud to radio-quiet (or radio-silent).  To explore how this conclusion would be biased by the large possibile parameter range inherent to TDFs, we compared our upper limits to currently available jet models, taking into account Doppler boosting and temporal evolution of the radio emission. 

We used our observations to constrain the jet model of \citet{vanVelzen11}. For a jet Lorentz factor of $\Gamma_j=5$, we can rule out the optimistic (``alway radio-loud'') scenario for four of the seven flares. The probability that the other three TDF candidates did launch such jets, but are not detected because Doppler boosting reduced the flux below two times the image rms is only 4\%. The hypothesis that all events hosted jets that only becomes radio-loud when the fallback rate drops below 2\% of the Eddington accretion rate (i.e., as observed in stellar mass black holes) is less constrained. Only for jets with $\Gamma_j<3$ this hypothesis is ruled out at 95\% CL. Our results are consistent with the recent radio observations of X-ray detected TDFs by \citet{Bower13}, which yielded two detections  (Table~\ref{tab:prev}), implying that 0--10\% of these flares launched relativistic jets (the radio-loud fraction in this sample could be zero because an AGN-origin for the radio emission of the two detected events is currently not ruled out).

We have also investigated the possibility that our sample of TDFs hosted a jet which is identical to Sw~1644+57, but oriented at a larger angle between the observer and the jet. Under the conservative assumption that jet Lorentz factor is constant ($\Gamma_j=2$), the estimated off-axis light curves of this relativistic TDF are inconsistent with the non-detection for four of the seven flares, for all possible observer angles. The hypothesis that all of the other TDFs hosted jets identical to Sw~1644+57 is ruled out at the 95\% confidence level.

Our results are not sensitive to our assumption that the time of disruption equals the time of the {\it Swift} trigger. If the hard X-rays of the jet are emitted only after ten times the fallback time ($\sim 1$~yr), the predicted off-axis flux is increased by just 50\%. A more serious caveat is that the off-axis light curves we used in this work are only valid for circumnuclear environments that are identical to the host of Sw~1644+57. This is not likely to be the case: the blue colors of optical/UV flares imply little optical extinction (i.e., reddening), while for Sw~1644+57 this extinction is much higher, $A_V = 3-5~{\rm mag}$ \citep{Bloom11}. Finally, we note that the black hole mass of  Sw~1644+57 may be a factor 5--10 smaller than the median black hole mass of thermal TDFs that we followed-up. The duration of the super-Eddington fallback rate of the flares in our sample may therefore be a year shorter, while the total jet energy could be factor 3 higher (for an accretion rate that is capped at the Eddington limit). A more sophisticated treatment of the off-axis light curves in the external model should take these differences into account, e.g., using radiative transfer onto the output of 2D hydrodynamical simulations \citep{van-Eerten11} for a range of black hole masses and environments. 

Definite proof that relativistic TDFs with evidence for jetted emission are an intrinsically different class can be obtained by radio transient surveys. For a disruption rate that is of the same of order as the rate of thermal TDFs, the areal density of radio transients like Sw~1644+57 (Eq. \ref{eq:snap}) almost exceeds the current upper limits. Near-future radio variability surveys, such as VAST \citep{Murphy12}, ThunderKAT, which is part of or MeerKAT \citep{Booth09}, or the LOFAR Transients Key Science Project \citep{Fender12}, will either detect tens to hundreds of TD jets per year, or conclude that the rate of stellar tidal disruptions with jets is lower than the rate of thermal TDFs. 

If a division between tidal disruptions with and without jets indeed exists, it presents a challenge to the idea that radio-loudness can be explained by state changes of accretion disk. Some authors have argued that the spin of the black hole is an important paramater in the production of stellar tidal disruption jets  \citep[e.g.,][]{Lei11, Krolik12, Stone12}:  TD jets may require rapidly spinning black holes, the presence of a pre-existing accretion disk, and/or the alignment of the disk angular momentum vector and the black hole spin vector. Observations of the emission from the accretion disk of tidal disruptions with jets will help to test these ideas.

\begin{acknowledgements}
SvV would like to thank the anonymous referee for the quick reply and useful comments. The VLA is operated by the National Radio Astronomy Observatory, a facility of the National Science Foundation operated under cooperative agreement by Associated Universities, Inc.
\end{acknowledgements}

\bibliography{general_desk}

\end{document}